\documentclass[useAMS,usenatbib]{mn2e}

\usepackage{graphicx} 
\usepackage{epsfig}
\usepackage{amsmath} 
\usepackage{txfonts}

\def\refe@jnl#1{{#1}}
\def\aj{\refe@jnl{Astron.~J.}}                  
\def\araa{\refe@jnl{Annu.~Rev.~Astron.~Astrophys.}}
\def\apj{\refe@jnl{Astrophys.~J.}}                 
\def\apjl{\refe@jnl{Astrophys.~J.~Lett.}}          
\def\aap{\refe@jnl{Astron.~Astrophys.}}            
\def\mnras{\refe@jnl{Mon.~Not.~R.~Astron.~Soc.}}   
\def\prd{\refe@jnl{Phys.~Rev.~D}}        
\def\fcp{\refe@jnl{Fund.~Cos.~Phys.}}  
\def\physrep{\refe@jnl{Phys.~Rep.}}   
\def\physlett{\refe@jnl{Phys.~Lett.}}

\title[Biases on the cosmological parameters and thermal SZ residuals]{Biases on the cosmological parameters and thermal Sunyaev--Zel'dovich residuals}
\author[N. Taburet et al.]{Nicolas Taburet\thanks{E-mail:
    nicolas.taburet@ias.u-psud.fr}, Nabila Aghanim,
  Marian Douspis and Mathieu Langer\\ Institut d'Astrophysique Spatiale, Universit\'e Paris Sud 11 and CNRS (UMR 8617), B\^at 121, 91405 ORSAY Cedex}
\begin{document}

\date{Accepted 2008 October 16. Received 2008 October 9; in original form 2008 July 16}

\pagerange{\pageref{firstpage}--\pageref{lastpage}} \pubyear{2009}

\maketitle
\label{firstpage}

\begin{abstract}
We examine the biases induced on cosmological parameters when the
presence of secondary anisotropies is not taken into account in Cosmic
Microwave Background analyses. We first develop an exact analytical
expression for computing the biases on parameters when any additive
signal is neglected in the analysis. We then apply it in the context
of the forthcoming \emph{Planck} experiment. For illustration, we
investigate the effect of the sole residual thermal
Sunyaev--Zel'dovich signal that remains after cluster
extraction. We find in particular that analyses neglecting the
  presence of this contribution introduce on the cosmological
  parameters $n_{\rm s}$ and $\tau$ biases, at least $\sim 6.5$ and
  $2.9$ times their one $\sigma$ confidence intervals. The
  $\Omega_{\rm b}$ parameter is also biased to a lesser extent.
\end{abstract}


\section{Introduction}

Future Cosmic Microwave Background (CMB) experiments, which
  are designed to be cosmic variance limited, will allow us to
determine the cosmological parameters with a relative
precision of the order of, or better than, one percent.  It will be
made possible in particular through the measurement of CMB temperature
and polarisation anisotropies with unequalled sensitivities, exquisite
angular resolution and optimal frequency coverage.  In this context,
additional contributions to the signal (galaxies, point sources,
secondaries arising from the interaction of CMB photons with matter
after decoupling, etc.)  can no more be neglected. More specifically,
a precise quantification of the biases on the parameters and of their
sources is now needed.

The study of biases in cosmology is receiving growing attention. In
the context of weak lensing, \citet{amara07} have derived a method
based on a Fisher matrix type analysis for quantifying systematic
biases.  In CMB analyses, \citet{Miller08} examined the biases
introduced by beam systematics for five upcoming experiments that will
measure the B-mode polarisation (\emph{Planck}, PolarBeaR, Spider,
Q/U Imaging Experiment (QUIET)+Clover and CMBPol).  Similarly, using \emph{Planck}
characteristics, previous studies have estimated the biases induced by
the contribution from patchy reionisation \citep{zahn05}. They
concluded that the biases, depending on the model of reionisation, can
be as high as a few in units of the one sigma error.  More recently,
\citet{serra08} have focused on the contribution from clustered IR
point sources and its effect on the cosmological parameters. Those two
studies used quite different approaches. While the
\citet{serra08} analysis was based on Monte Carlo Markov Chains
(MCMC), \citeauthor{zahn05} used an approximate analytic computation
of the biases \citep[see also][]{huterer06}.

In this study, we present an analytical derivation of the biases on
the cosmological parameters that goes beyond the aforementioned
approximations.  It is a method valid when the primary signal and
secondary contribution (astrophysical or systematic) are additive and
it is exact as it can be applied even when the secondary signal is
dominant over the primary. The method presented here is applied to the
estimate of the cosmological parameters with the CMB power spectrum.
We furthermore focus on one single source of additional anisotropies:
those associated with undetected clusters. The Sunyaev--Zel'dovich
(SZ) effect of galaxy clusters \citep{SunyaevZeldovich72} is indeed
the major source of secondary temperature anisotropies \citep[][ and
  references therein]{Aghanimrevue08}.  The SZ effect is two-fold: the
thermal SZ due to the inverse Compton scattering of photons off the
hot electrons in the intra-cluster gas \citep[e.g.][]{Rephaeli95,
  Itoh98}, and the kinetic effect due to the Doppler shift caused by
the proper motion of the clusters in the CMB reference frame.  The
upcoming large multi-frequency surveys will be able to detect and
extract galaxy clusters using their specific thermal SZ spectral
signature. We nevertheless expect some level of residual SZ
contribution from undetected clusters in the temperature anisotropy
maps. Such a residual signal might be the cause of the excess of power
measured by small scale CMB experiments like BIMA (Berkeley Illinois Maryland Association Array), CBI, ACBAR
\citep{bima, cbi, acbar, cbi2}. The excess could be also due to
unremoved point sources \citep[][]{toffolatti05,Marian06}.

Our article is organised as follows, we present in Section 2 the
method to calculate the biases on cosmological parameters. We then apply
our method to estimate the biases induced by the SZ residuals. We
present our results in Section 3 and discuss them in the following
section. Finally, we conclude in Section 5.  Throughout the article,
we assume a flat $\Lambda$ cold dark matter ($\Lambda$CDM) cosmological model and use for the
reference model the \emph{Wilkinson Microwave Anisotropy Probe5} (WMAP5) cosmological parameters \citep{wmap5},
unless otherwise stated.


\section{Derivation of the biases}

In order to forecast the capability of instruments in terms of
parameter estimation, we commonly use the Fisher approach as it
provides us with the achievable precision on the parameters as a function of the
instrumental characteristics (noise, beam, etc.) However, predicting
the precision is often not enough, and one should, in addition, estimate
how sensitive the measurements are to any systematic or additional
contribution to the signal. The aim of our study is to propose an
analytical calculation of the bias introduced on the estimated
parameters by the presence of an additional, non primary signal. We
consider the latter to be any signal that is added, or subtracted, to form
the observed signal used for parameter estimation. The method
presented here is thus general and can be applied not only to astrophysical
contaminating signals (e.g. foregrounds) but also to instrumental
systematics when they are additive. We develop our method in the
framework of cosmological parameter estimation using CMB angular power
spectra and we apply it to the case of future \emph{Planck} observations in
Section \ref{sect:application}.

Let us assume that the signal ($C_{\ell}^{\rm D}$) used to estimate
cosmological parameters is the sum of a primary signal [$C_\ell^{\rm
  CMB}(\boldsymbol{\hat{\theta}})$] that depends on cosmological parameters
$\boldsymbol{\hat{\theta}}$ and an additional signal ($C_\ell^{\rm add}$)
that may, or may not, contain cosmological information. If such an
additional signal is taken into account in the parameter estimation
analysis (Case 1) then one recovers the ``true'' cosmological
parameters, ${_1}\boldsymbol{\hat{\theta}}$. If, on the contrary, one 
uses the total signal but assumes that it is made of primary only (Case 2),
one obtains a set of biased cosmological parameters,
${_2}\boldsymbol{\hat{\theta}}$. In the following, we derive an analytical
expression of this bias, $\boldsymbol{\hat b}={_2}\boldsymbol{\hat\theta}-{_1}\boldsymbol{\hat\theta}$.

We consider that the errors associated with the ``data" $C_\ell^D$ are
distributed according to a Gaussian law. The likelihood function is
then written as $\mathcal{L}=e^{-\chi^2/2}$.  For Case 1, the
$\chi_1^2$ is :
\begin{equation}
\label{eq:chi2_1}
\chi_1^2(\boldsymbol{\hat\theta})=\sum_{\ell,X,Y}{\rm
  cov}_\ell^{-1}(C_\ell^XC_\ell^Y)\left[C_\ell^{{\rm D}X}-C_\ell^{X \rm
  mod'}(\boldsymbol{\hat\theta})\right]\times\left[C_\ell^{{\rm D}Y}-C_\ell^{Y \rm
  mod'}(\boldsymbol{\hat\theta})\right]
\end{equation}
where $C_\ell^{X \rm mod'}=C_\ell^{X \rm CMB}+C_\ell^{X {\rm add}}$.
For Case 2, the $\chi_2^2$ is given by:
\begin{equation}
\label{eq:chi2_2}
\chi_2^2(\boldsymbol{\hat\theta})=\sum_{\ell,X,Y}{\rm
  cov}_\ell^{-1}(C_\ell^XC_\ell^Y)\left[C_\ell^{{\rm D}X}-C_\ell^{X \rm
  mod}(\boldsymbol{\hat\theta})\right]\times\left[C_\ell^{{\rm D}Y}-C_\ell^{Y \rm mod}(\boldsymbol{\hat\theta})\right]
\end{equation}
where $C_\ell^{X \rm mod}=C_\ell^{X \rm CMB}$ only, and $X,Y=$
  TT, EE, TE.  We consider the temperature and E-mode
  auto-correlations, TT and EE, and the cross-correlation, TE, for
  which the coefficients of the covariance matrix are
  cov$_\ell(C_\ell^XC_\ell^Y)$ \citep[see, e.g.][]{ZaldSelj97}.

Assuming the "data" represent the sum of primordial and
additional signals, $\left\langle C_\ell^{\rm D}\right\rangle
=C_\ell^{\rmn{CMB}}({_1}\boldsymbol{\hat{\theta}})+C_\ell^{\rmn{add}}$, the set of
parameters ${_1}\boldsymbol{\hat\theta}$ minimises $\langle\chi_1^2\rangle$ and
the parameter set ${_2}\boldsymbol{\hat\theta}$ minimises $\langle\chi_2^2\rangle$:
\begin{equation}
\label{eq:dernulle}
\forall i, \left.\frac{\partial\left\langle \chi^2_1\right\rangle
}{\partial\theta_i}\right|_{\boldsymbol{\hat\theta}={_1}\boldsymbol{\hat\theta}}=\left.\frac{\partial\left\langle
  \chi^2_2\right\rangle
}{\partial\theta_i}\right|_{\boldsymbol{\hat\theta}={_2}\boldsymbol{\hat\theta}}=0.
\end{equation}
An ensemble average of a second order approximation of equation
(\ref{eq:chi2_2}) gives:
\begin{equation}
\label{eq:DL2moy}
\left\langle \chi^2_2({_2}\boldsymbol{\hat\theta})\right\rangle =\left\langle
\chi^2_2({_1}\boldsymbol{\hat\theta})\right\rangle +\sum_ib_i\left\langle
\left.\frac{\partial\chi^2_2}{\partial\theta_i}\right|_{\boldsymbol{\hat\theta} =
  {_1}\boldsymbol{\hat\theta}}\right\rangle +\frac{1}{2}\sum_{ij}b_ib_j\left\langle
\left.\frac{\partial^2\chi^2_2}{\partial\theta_i\partial\theta_j}\right|_{\boldsymbol{\hat\theta}={_1}\boldsymbol{\hat\theta}}\right\rangle. 
\end{equation}
Using equations (\ref{eq:dernulle}) and (\ref{eq:DL2moy}), we obtain:
\begin{equation}
\label{eq:forbias}
\forall i, \left\langle \left.\frac{\partial\chi^2_2}{\partial\theta_i}\right|_{\boldsymbol{\hat\theta}={_1}\boldsymbol{\hat\theta}}\right\rangle 
=-\sum_{j}b_j\left\langle \left.\frac{\partial^2\chi^2_2}{\partial\theta_i\partial\theta_j}
\right|_{\boldsymbol{\hat\theta}={_1}\boldsymbol{\hat\theta}}\right\rangle
\end{equation}
that can be written in the compact form
\begin{equation}
\boldsymbol{V} = \mathbf{G}\,\boldsymbol{\hat{b}}
\end{equation}
wherefrom the bias vector follows simply,
\begin{equation}\label{eq:biases}
\boldsymbol{\hat{b}} = \mathbf{G}^{-1}\, \boldsymbol{V}.
\end{equation}
From equation (\ref{eq:chi2_2}), in the context of CMB angular power spectra measurements,
\begin{eqnarray}
\label{eq:Fisherforbiases}
G_{ij} & = &\sum_{\ell,X,Y}{\rm
  cov}_\ell^{-1}(C_\ell^XC_\ell^Y)\left(\left.\frac{\partial
    C_\ell^{X\rm
      mod}}{\partial\theta_i}\right|_{\boldsymbol{\hat\theta}=_1\boldsymbol{\hat\theta}}\left.\frac{\partial
    C_\ell^{Y\rm
      mod}}{\partial\theta_j}\right|_{\boldsymbol{\hat\theta}=_1\boldsymbol{\hat\theta}}\right.\nonumber\\ &
  & +\left.\frac{\partial C_\ell^{X\rm
      mod}}{\partial\theta_j}\right|_{\boldsymbol{\hat\theta}=_1\boldsymbol{\hat\theta}}\left.\frac{\partial
    C_\ell^{Y\rm
      mod}}{\partial\theta_i}\right|_{\boldsymbol{\hat\theta}=_1\boldsymbol{\hat\theta}}\nonumber\\ &
  & -\left.C_\ell^{X{\rm add}}\left.\frac{\partial^2 C_\ell^{Y\rm
      mod}}{\partial\theta_i\partial\theta_j}\right|_{\boldsymbol{\hat\theta}=_1\boldsymbol{\hat\theta}}-C_\ell^{Y{\rm
      add}}\left.\frac{\partial^2 C_\ell^{X\rm
      mod}}{\partial\theta_i\partial\theta_j}\right|_{\boldsymbol{\hat\theta}=_1\boldsymbol{\hat\theta}}\right)
\end{eqnarray}
and 
\begin{eqnarray}\label{eq:vector}
V_i & = & \sum_{\ell,X,Y}{\rm
  cov}_\ell^{-1}(C_\ell^XC_\ell^Y)\left(C_\ell^{Y{\rm
      add}}\left.\frac{\partial C_\ell^{X\rm
      mod}}{\partial\theta_i}\right|_{\boldsymbol{\hat\theta}=_1\boldsymbol{\hat\theta}}\right.\nonumber\\ &
  & \left.+C_\ell^{X{\rm add}}\left.\frac{\partial C_\ell^{Y\rm
      mod}}{\partial\theta_i}\right|_{\boldsymbol{\hat\theta}=_1\boldsymbol{\hat\theta}}\right).
\end{eqnarray}
Equation (\ref{eq:forbias}) thus allows us to calculate the bias on
a parameter $\theta_i$ as a function of the additional signal and of the
first and second derivatives of the primary signal. The computational advantage 
is that any additive contribution can be readily inserted without the need of 
a re-computation of the derivatives of the primary signal.

In the case of CMB (temperature and polarisation) observations, when
an additional signal is ignored, the biases induced on cosmological
parameters are obtained from equations (\ref{eq:biases}--\ref{eq:vector}). The derived formula
(equation \ref{eq:forbias}) can, however, be used in many other cases (matter
power spectrum, weak lensing power spectrum, etc.), as it accounts
quite generally for additive contributions or systematics.
Furthermore, it is the most general expression for the biases (for
additive signals) as it applies to cases where the secondary signal is
dominant over the primary.  In contrast, previous studies
\citep[e.g. ][]{zahn05,huterer06} of the bias on cosmological
parameters used approximations applicable only when the additional
signal is negligible with respect to the primary signal.

The biases on the investigated parameters become relevant only if
they are larger than the expected confidence intervals. The latter
can be computed through a Fisher matrix analysis. The 68.3 per cent
confidence interval on one parameter (the others being known) is given
by \citep[][]{numrec}:
\begin{equation}
\delta\theta_i=\sqrt{F^{-1}_{ij}}, 
\end{equation}
where the matrix coefficients are:
\begin{equation}
F_{ij}=\sum_\ell\sum_{X,Y}{\rm cov}_\ell^{-1}(C_\ell^XC_\ell^Y)\frac{\partial
C_\ell^X}{\partial\theta_i}\frac{\partial C_\ell^Y}{\partial\theta_j}.
\end{equation}
The numerical values for the instrumental noise and the beam used
  to compute the covariance matrix can be found in the \emph{The
    Scientific Programme of Planck} \citep[also known as the
    \emph{Planck Blue Book}, ][]{PlanckBB}.


\section{Computation of the thermal SZ residual}
\label{sect:application}

The \emph{Planck} satellite will measure CMB anisotropies with an
unprecedented precision, in temperature and polarisation, over the
full sky and from the largest scales down to five
arcminutes. Foregrounds and secondary anisotropies are expected to
contribute to the signal. Taking advantage of the multi-frequency
observation, cleaning algorithms are developed to disentangle the
primary signal from contaminants.  On small scales, the SZ effect from
galaxy clusters will be one of the major secondary contribution to the
signal. Fortunately, the characteristic spectral signature of the
thermal SZ effect makes it easily detectable. On the one
hand, one can remove some of the thermal SZ signal from the
CMB maps in order to recover the best primary CMB angular power
spectrum, and on the other hand a cluster catalogue can be built
\citep[e.g.][]{bjorn2006}.  Nevertheless, some level of residual,
unresolved SZ signal will remain in the temperature maps. We thus
apply the method described in the previous section to estimate the
bias induced by that SZ residual signal in the CMB maps on the six
`standard' cosmological parameters.

We now compute the additional contribution to the CMB signal that
enters equation (\ref{eq:biases}). Here, it is simply the power
spectrum of the residual SZ signal that remains after cluster
extraction. We focus on the Poisson contribution to the SZ angular
power spectrum following \citet[][]{komatsu02} and assume that the
contribution from correlated halos is negligible \citep[valid for
  $\ell>300$,][]{komatsu99}:
\begin{equation}
\label{eq:SZ_powspec}
C_\ell=f^2(x)\int_0^{z_\rmn{max}}dz\frac{dV_\rmn{c}}{dzd\Omega}\int_{M_{\rm
    min}}^{M_{\rm max}}dM\frac{dn(M,z)}{dM}\left|\tilde{y_\ell}(M,z)\right|^2,
\end{equation}
where $\frac{dV_\rmn{c}}{dzd\Omega}$ is the comoving volume per unit
redshift and solid angle and $n(M,z)dM\frac{dV_\rmn{c}}{dzd\Omega}$ is
the probability of having a galaxy cluster of mass $M$ at a redshift
$z$ in the direction $d\Omega$.  In the present study, we use the
\citet[][]{ShethTormen} mass function $n(M,z)$.  The SZ frequency
dependence is encoded in the function $f(x)=x\frac{e^x+1}{e^x-1}-4$
where $x= h_\rmn{Pl}\nu/k_{\rm B}T_{\rm e}$ is the dimensionless
frequency, and $\tilde{y}_\ell=\tilde{y}_\ell(M,z)$ is the
two-dimensional Fourier transform on the sphere of the 3D radial
profile of the Compton $y$-parameter of individual clusters,
\begin{equation}
\tilde{y}_\ell=\frac{4\pi}{D_{\rm A}^2}\int_0^\infty
y_{\rm 3D}(r)\frac{\sin(\ell r/D_{\rm A})}{\ell r/D_{\rm A}}r^2dr,
\end{equation}
with $y_{\rm 3D}(r)= \sigma_{\rm T}\frac{k_{\rm B}T_{\rm e}(r)}{m_{\rm
    e}c^2}n_{\rm e}(r)$ and $D_{\rm A} = D_{\rm A}(z)$ the proper
angular-diameter distance.  To model the SZ signal from individual
clusters, we assume that the electron density $n_{\rm e}(r)$ within
the cluster virial radius $r_\rmn{vir}$ follows a spherical isothermal
$\beta$--profile \citep[][]{cavaliere76} with core radius
$r_c=0.1\,r_\rmn{vir}$ and $\beta=2/3$ for all clusters for
simplicity. The isothermal temperature of the intra-cluster gas
$T_{\rm e}$ is set equal to the virial temperature.

In order to evaluate equation (\ref{eq:SZ_powspec}), we choose
$z_\rmn{max}=7$ and $M_{\rm min}=10^{13}$M$_\odot$. The angular power
spectrum of the residual SZ is the contribution from undetected
clusters. It is computed by setting $M_{\rm max}=M_\rmn{lim}(z)$, the
cluster detection limit.
\begin{figure*}
\begin{minipage}{.49\linewidth}
\centering
\includegraphics[width=8cm]{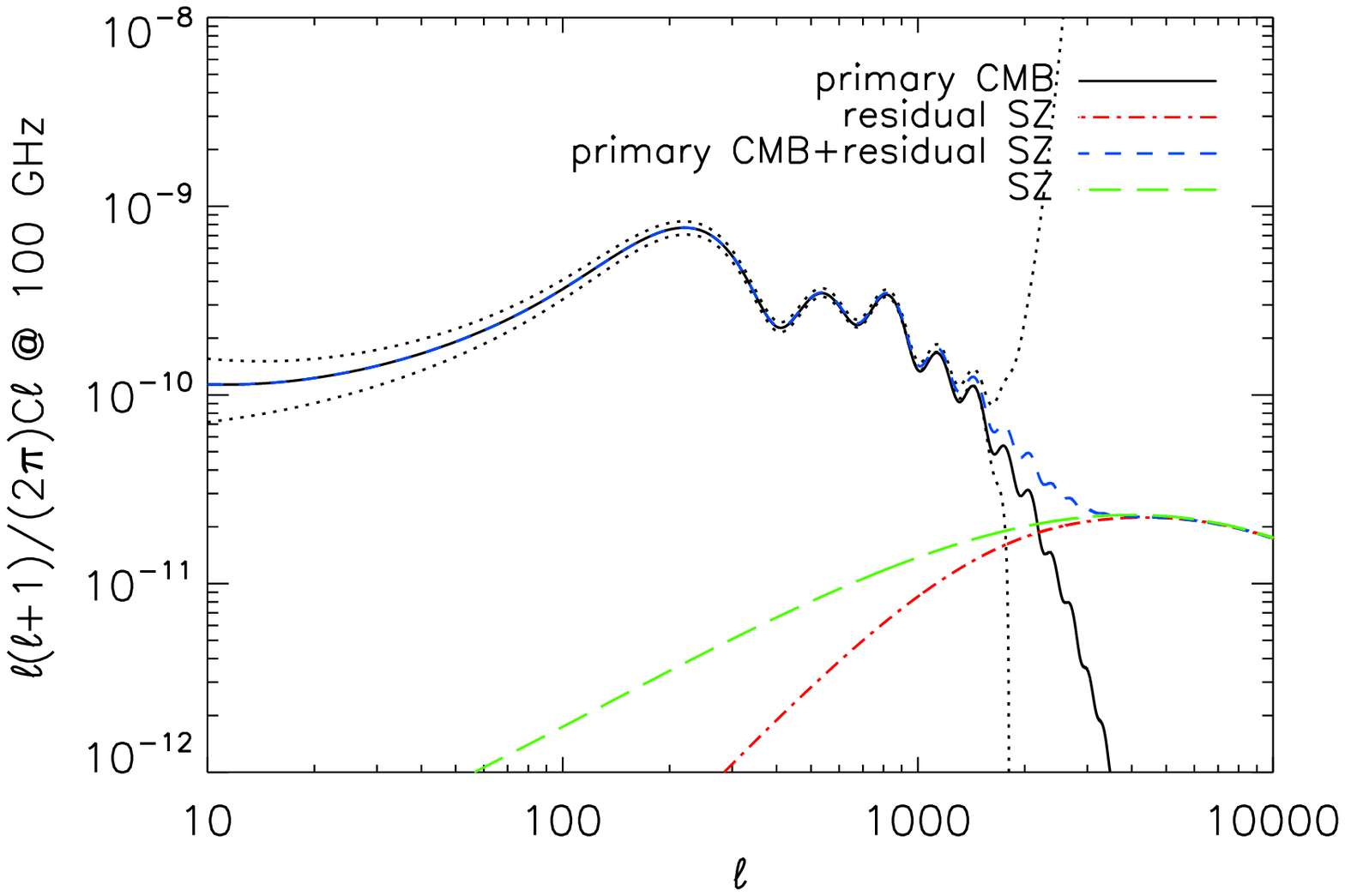}
\end{minipage}
\begin{minipage}{.49\linewidth}
\centering
\includegraphics[width=8cm]{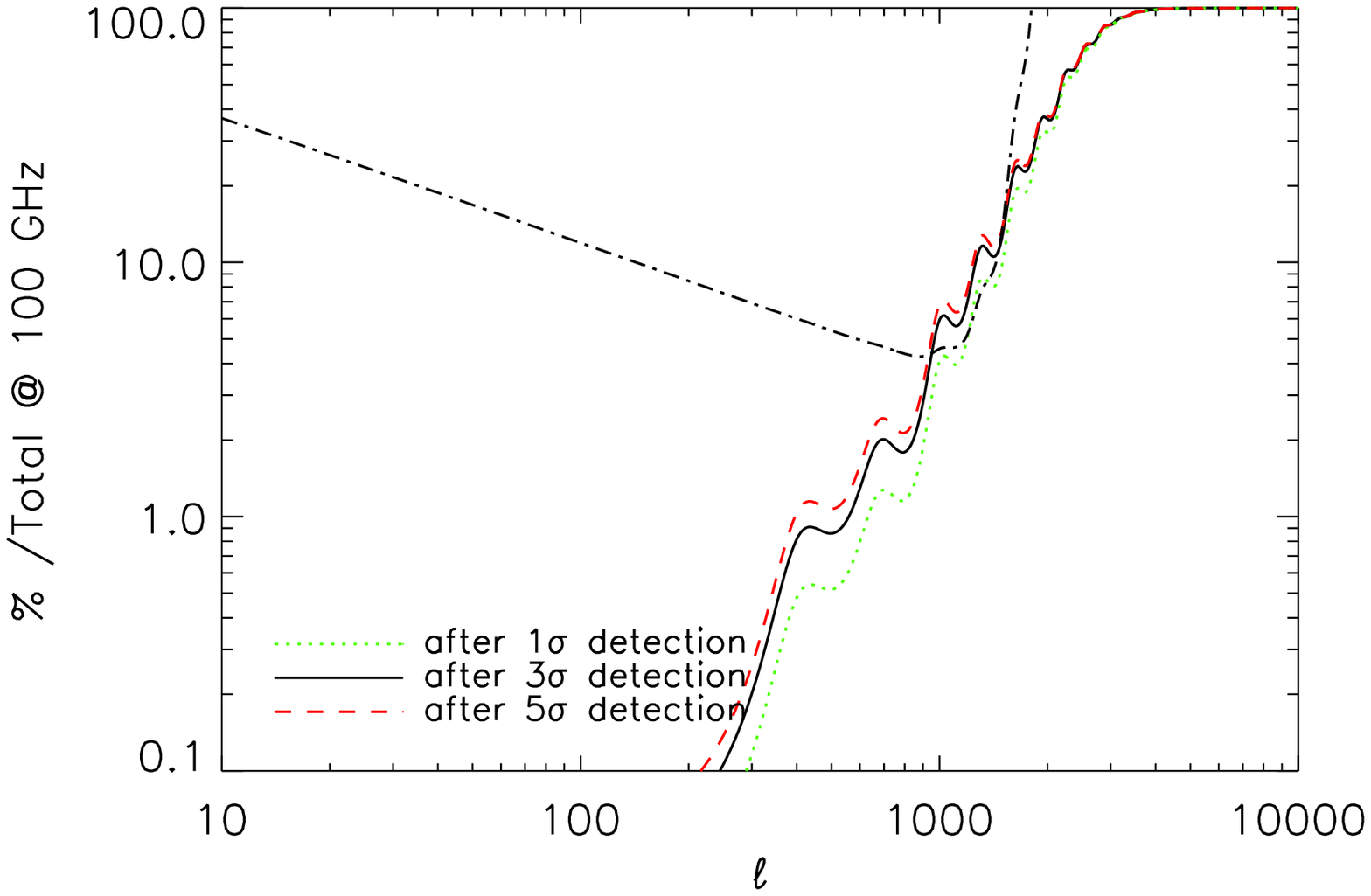}
\end{minipage}
  \caption{Left-hand panel: At 100 GHz, primary CMB (solid line), SZ
    angular power spectrum of the whole cluster population
    (long-dashed green line), residual SZ spectrum after the
    extraction of clusters above $3\sigma_Y$ simultaneously at 100,
    143 and 353 GHz (dot-dashed red line), primordial CMB + residual
    SZ spectrum (dashed blue line). The black dotted lines represent
    the 1$\sigma$ error bars for the 100 GHz \emph{Planck}
    channel. Right-hand panel: Contribution of the residual SZ power to the
    total signal after $3\sigma_Y$ cluster detections (solid black
    line), 1$\sigma_Y$ (green dotted line) and 5$\sigma_Y$ (dashed red
    line). The envelop represents the instrumental
      sensitivity and the cosmic variance. }
\label{fig:CMB+SZ}
\end{figure*}
The selection function determines the limiting mass $M_\rmn{lim}(z)$
as a function of redshift for a cluster to be detected by a given
instrument characterised by its beam, sensitivity and frequency
coverage.  We compute the selection function similarly to
\citet[][]{Bartelmann01}.  A galaxy cluster is detected if at the same
time its beam-convolved Compton parameter $\bar y(\theta)$ emerges
from the confusion noise, and if its integrated signal is above the
instrument sensitivity. This translates into the following condition
on the flux variation:
\begin{equation} 
\label{flux_fin}
\overline{\Delta F(x)}=g(x)\,\rmn{I}_0\,\int\bar y(\theta)d\Omega\geq
 \lambda\,g(x)\,\rmn{I}_0\bar{Y}_{\rm lim},
\end{equation}
where $g(x) = \frac{x^4e^x f(x)}{(e^x-1)^2}$ and I$_0=2\frac{(k_{\rm
    B}T_\rmn{CMB})^3}{(h_\rmn{Pl}c)^2}$.  The integral is calculated
over the lines of sight for which $\bar y(\theta)\geq 3\,\Delta y_{\rm
  bg}$, and the sensitivity limit $\bar{Y}_{\rm lim}$ is derived from
the antenna temperature sensitivity of the instrument.

The background SZ signal responsible for the confusion is computed
assuming a Poisson distribution of clusters with masses between
$10^{13}$M$_\odot$ and $5\times 10^{16}$M$_\odot$:
\begin{equation}
\label{rms_ybg}
\Delta y_{\rm bg}=\sqrt{\int dz\frac{dV_\rmn{c}}{dzd\Omega}\int
  dM\, n(M,z)\left(\int d\Omega\, y(\theta,M,z)\right)^2}.
\end{equation}
Finally, the flux in equation (\ref{flux_fin}) is integrated over a
top hat function to account for the frequency response of the
instrument, and the integer $\lambda$ is the detection threshold in
terms of the instrumental noise $\sigma_Y$.

In order to compute the SZ residual temperature power spectrum that
enters equations (\ref{eq:biases}) and (\ref{eq:Fisherforbiases}), we
derive an idealised cluster selection function for \emph{Planck} (not
accounting for the effects of foregrounds and point sources)
considering only the channels relevant for the SZ measurement, namely
100, 143 and 353 GHz (217 GHz is the frequency for which the thermal
SZ is null). We consider that an SZ cluster is detected only if the
selection criterion (\ref{flux_fin}) is satisfied in all three
frequency channels simultaneously.

\section{Results} 
\label{sect:results}

We now illustrate the application of our formalism by taking the
future \emph{Planck} satellite as an example of high sensitivity high
resolution CMB experiment. The instrumental characteristics that we
used can be found in the \emph{Planck Blue Book} \citep[][]{PlanckBB}.  We quantify
simultaneously the biases induced on the cosmological parameters
$\Omega_\Lambda$, $\Omega_{\rm b}$, $H_0$, $n_{\rm s}$, $\sigma_8$ and
$\tau$ by the residual SZ contribution at 100 GHz.  As a reference
model, we take the cosmological parameters obtained by the \emph{WMAP} team
\citep[][]{wmap5} and the corresponding $C_\ell^{\rm CMB}+C_\ell^{{\rm
    res}}$, where $C_\ell^{{\rm res}}$ is the residual SZ signal.

Undetected clusters contribute as a
residual signal at high multipoles where they dominate over the
primordial CMB for $\ell$ higher than 2000 approximately (Figure
\ref{fig:CMB+SZ}, left panel).  At $\ell>1000$, we show on the right
panel of figure \ref{fig:CMB+SZ} that the residual SZ contribution
after extracting clusters above $3\sigma_Y$ already represents more than
10 per cent of the total signal and exceeds the instrumental noise.

\begin{figure*}
\includegraphics[width=16cm]{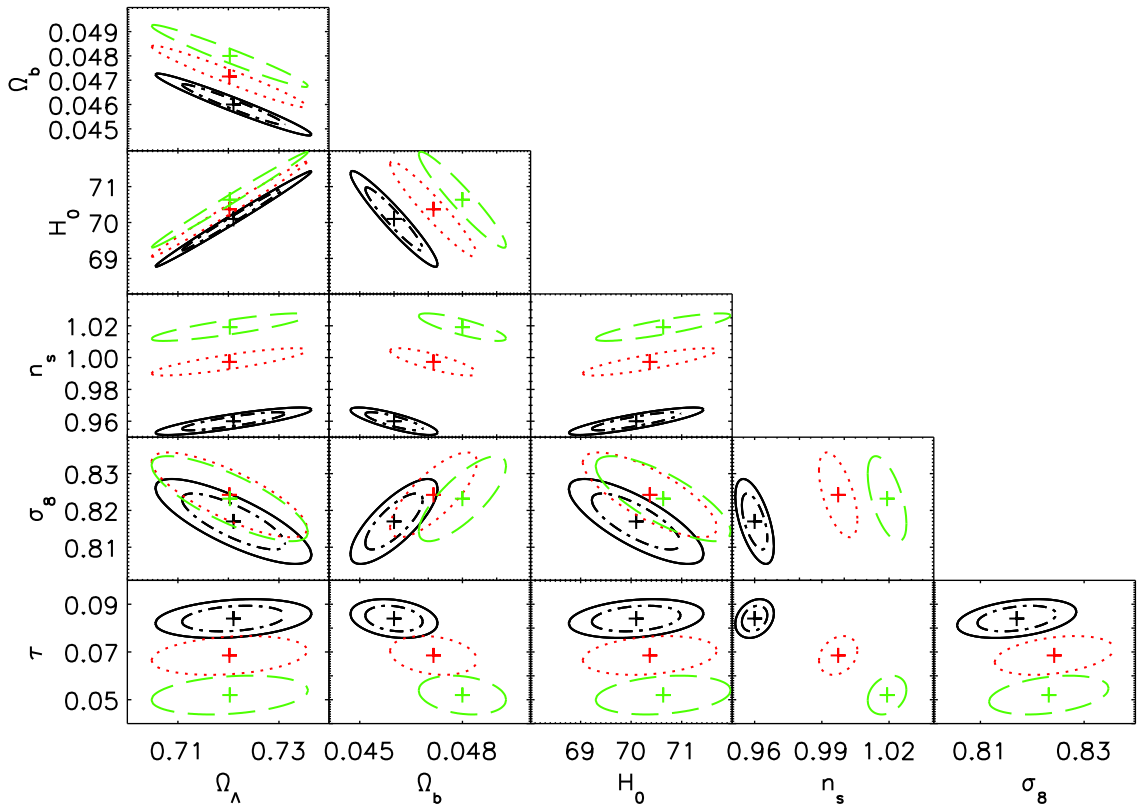}
  \caption{68.3 per cent joint confidence regions for $\Omega_\Lambda$,
    $\Omega_{\rm b}$, $H_0$, $n_{\rm s}$, $\sigma_8$ and $\tau$ (solid
    line) obtained with the expected \emph{Planck} TT, TE and EE
    spectra computed for the reference cosmological model.  The dotted
    (red) and long dashed (green) shifted ellipses represent the
    68.3 per cent joint confidence regions around the biased parameters if
    respectively the 1$\sigma_Y$ and 5$\sigma_Y$ residual SZ signal is
    not taken into account.}
\label{fig:bias_6params}
\end{figure*}

In equations (\ref{eq:biases}) and (\ref{eq:Fisherforbiases}), we do not
consider polarised residuals at high $\ell$s since
the polarisation induced by galaxy clusters is negligible compared to
the primary CMB polarisation at those scales \citep[e.g.][]{Liu05}. Therefore
\begin{equation}
C_\ell^{X{\rm res}}(\hat\theta)=\left\{\begin{array}{cl}
C_\ell^{{\rm TT res}}(\hat\theta)& \text{if } X={\rm TT}\\
0 & {\rm otherwise.} 
\end{array}\right.\\
\end{equation}

We show the induced biases in figure \ref{fig:bias_6params}.  The
black solid lines represent the 68.3 per cent joint confidence level ellipses
on the parameters $\Omega_\Lambda$, $\Omega_{\rm b}$, $H_0$, $n_{\rm
  s}$, $\sigma_8$ and $\tau$ using the TT, TE and EE power spectra
centered on the reference model.  As shown in the figure, the
expected constraints on cosmological parameters are at a few percent
precision. We show, in addition, the biased values of the parameters
due to the residual SZ signal.  We consider different cluster
detection limits (1, 3, and 5$\sigma_Y$) and thus obtain various
levels of residuals in the signal.  The results for all parameters are
summarised in Table \ref{tab:bias}. In figure \ref{fig:bias_6params},
the dotted red and long dashed green ellipses, shifted with
  respect to those in solid black, represent the 68.3 per cent joint
confidence regions if the SZ residual signal that is obtained when
clusters are extracted above 1 or 5$\sigma_Y$ is not modelled
properly.

\begin{figure}
\includegraphics[width=8cm]{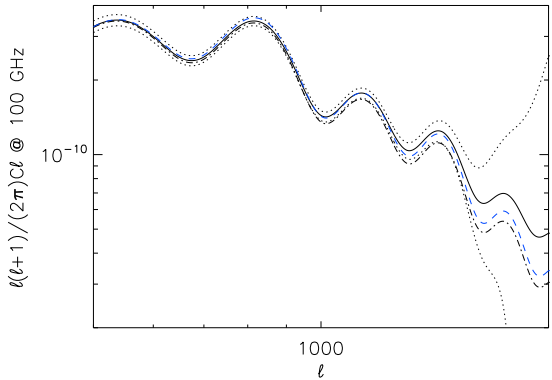}
  \caption{Solid line (reference model) : Primary CMB + SZ residuals
    calculated using the reference model. Dot-dashed line : Primary
    CMB calculated using the reference model. Blue dashed line :
    Primary CMB calculated with the biased parameters. The dotted
    lines are the envelopes representing the instrumental sensitivity
    around the reference model (CMB+residual SZ).}
\label{fig:curves}
\end{figure}

\begin{figure}
\includegraphics[width=8cm]{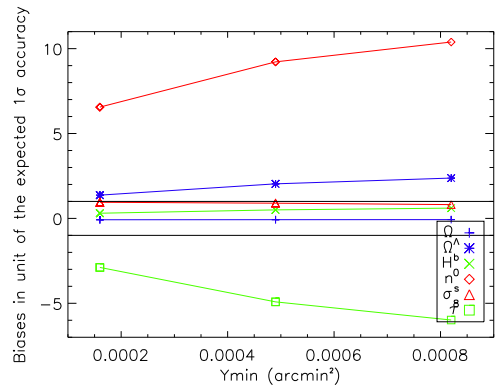}
  \caption{Biases, in units of the expected $1\sigma$ accuracy, induced
    by the SZ residuals as a function of the SZ cluster detection
    threshold for the 6 parameters case. $\Omega_\Lambda=$ blue
    pluses, $\Omega_{\rm b}=$ blue stars, $H_0=$ green x, $n_{\rm s}=$ red
    diamonds, $\sigma_8=$ red triangles and $\tau=$ green squares.}
\label{fig:biasesgraphik}
\end{figure}

First, and unsurprisingly, the biases induced on
$\Omega_\Lambda$ and $H_0$ are negligible. At most, i.e. for
a large residual contribution, they reach roughly 0.1 and 0.6
in units of the 1$\sigma$ errors on the parameters. Then, as expected,
$\sigma_8$, and to a larger extent $n_{\rm s}$, $\Omega_{\rm
    b}$ and $\tau$, are on the contrary significantly affected by the
  residual SZ signal. Those parameters are indeed, by nature, the
most sensitive to an excess of power at small scales; additionally
they are degenerate.  An excess of power at high $\ell$, due to SZ
residuals, can be accounted for by over-estimating the value of
$\sigma_8$. Since the amplitude of the CMB power spectrum
  strongly depends on $\sigma_8$, a relatively small variation of
  $\sigma_8$ is enough to fit the power excess. As a result, the bias
  on this parameter is rather small.  The excess of power at small
scales decreases the ratio between the amplitudes of the forth and
fifth CMB peaks and slightly shifts them towards higher $\ell$ values
(Fig. \ref{fig:curves}) mimicking the effects induced by an increase
of $\Omega_{\rm b}$. We find that the bias is roughly
1.4 and 2.4 in units of the error on $\Omega_{\rm b}$, for
the 1 and 5$\sigma_Y$ cases respectively.  Higher $\Omega_{\rm b}$ and
$\sigma_8$ values imply a higher amplitude of the primary CMB spectrum
at all multipoles. However, the SZ residual signal contributes
significantly to the CMB spectrum only at multipoles higher than 500,
leaving lower multipoles almost unaffected. The effects at low
multipoles of simultaneously higher $\Omega_{\rm b}$ and $\sigma_8$,
namely an overall rise of the power, are compensated by an increase of
the spectral index $n_{\rm s}$, which raises (lowers) the power at
multipoles larger (smaller) than $\sim 1200$. They are also
  compensated by a decrease of the optical depth $\tau$, that reduces
  the CMB power.  As a consequence, fitting data containing
primary CMB and SZ secondary residual with a pure primary CMB spectrum
induces quite an important bias on $n_{\rm s}$ and $\tau$. The
  bias on $\tau$ is between 2.8 and 6 times the expected precision for
  cluster detection limits 1 and 5$\sigma_Y$, respectively, and the
  bias on $n_{\rm s}$ is between 6.5 and 10.4 times the expected
  precision for 1 and 5$\sigma_Y$ cluster thresholds. All these biases are summarized in Fig. \ref{fig:biasesgraphik}
as functions of the SZ cluster detection threshold.

\begin{table*}
\label{tab:biases}
 \centering
\caption{Biases induced on cosmological parameters when the SZ signal
  due to undetected clusters is ignored.}
  \begin{tabular}{@{}cccccccc@{}}
  \hline \hline detection & detection threshold & \multicolumn{6}{c}{biases
    expressed in units of the expected precision}\\ 
 & $\bar{Y}_{\rm lim}$ (arcmin$^2$) &
  $\Omega_\Lambda$ & $\Omega_{\rm b}$ & $H_0$ & $n_{\rm s}$ & $\sigma_8$ & $\tau$ \\ 
\hline 
1$\sigma_Y$ & $1.6\times 10^{-4}$ & -0.078 & 1.36 & 0.3 & 6.49 & 0.92 & -2.78 \\ 
\hline 
3$\sigma_Y$ & $4.9\times 10^{-4}$ & -0.073 & 2.02 & 0.49 & 9.12 & 0.89 & -4.81\\ 
\hline
5$\sigma_Y$ & $8.2\times 10^{-4}$ & -0.07 & 2.36 & 0.6 & 10.35 & 0.78 & -5.96\\ 
\hline 
\hline
\end{tabular}
\label{tab:bias}
\end{table*}

\section{Discussion}
\label{sec_bias}

We show that an unremoved thermal SZ contribution in future high
sensitivity high resolution CMB experiments, measuring both
temperature and polarisation, introduces excess of power at small
angular scales. This in turn induces biases on the cosmological
parameters associated with inflation, the spectral index $n_{\rm s}$
and, to a lower extent, the normalisation $\sigma_8$, as well
as on the density of baryons $\Omega_{\rm b}$ and the optical
  depth $\tau$.  The other two parameters of the standard
model remain essentially unaffected.

The amplitude of the biases strongly depends on the amplitude and the
detailed shape of the residual power spectrum. In order to estimate
the residual thermal SZ power spectrum, we used a theoretical SZ
cluster selection function built from the expected capabilities of
\emph{Planck} and not accounting for foregrounds and point sources.
The detected clusters are expected to have masses of a few
$10^{14}$M$_\odot$ at redshifts $0.4<z<0.7$.  Although, the
theoretical selection function does not coincide totally with that
obtained from simulations (false detections, confusion with point
sources, etc., J.-B. Melin, private communication), it gives us a
reasonable estimate of the residual thermal SZ power spectrum used in
our analysis.

It is important to bear in mind that other astrophysical contributions
need to be taken into account in future CMB experiments observing at
small scales. In that respect, we have also
estimated the biases induced on the cosmological parameters by the
kinetic SZ anisotropies of the whole cluster population. Such a signal
cannot be separated out from the primary signal as they both share the
same spectral signature. The kinetic SZ signal, although one order of
magnitude smaller, will dominate over the thermal SZ at 217~GHz where
the latter vanishes.  We found that the bias on the spectral
  index $n_{\rm s}$ is less than 2$\sigma$, whereas the other
  parameters are essentially unaltered by the additional
  contribution. The channel at 217~GHz could seem more appropriate
for the determination of cosmological parameters. However, one needs
to consider additionally the contribution from extra-galactic point
sources, radio or IR galaxies and their clustering. A recent study by
\citet{serra08} has extended the work of \citet{Marian06} by focusing
on IR galaxy clustering with WMAP and ACBAR data, as well as with
\emph{Planck}. They showed (at 143~GHz) that the expected biases on
$\Omega_{\rm b}$, $n_{\rm s}$ and $\sigma_8$ are of the order of 2.

We have also applied our method to estimate possible biases on the
cosmological parameters with WMAP. Due to its frequency coverage ($\nu
< 94$ GHz) we can safely consider that the thermal SZ residual signal
in the data is constituted of the contribution from all the clusters
(none of them being actually detected and extracted). The angular
resolution of \emph{WMAP} is limited to $\ell < 1000$. In this range, the
contribution of the SZ signal to the CMB power does not exceed
10 per cent. As a result, although the contamination by thermal SZ is maximum
in the \emph{WMAP} case, we found that the derived cosmological parameters
are unbiased.

Given the possibility that large biases could arise due to various
systematics, it is important to find a way to reduce their impact.
First, one could consider only the multipoles for which the primary
contribution to the TT signal is dominant, i.e. typically $\ell <
1000$. This was done by \citet{zahn05} who suggested to take into
account the full polarisation power spectra but to use the TT power
spectrum only out to $\ell\sim 1000$.  We have tested such an
  approach on our issue: the SZ residuals. We find that truncating the
  measurements above $\ell= 1000$ artificially increases the
  error-bars on $\Omega_\Lambda$, $\Omega_{\rm b}$ and $H_0$ by
  approximately 10 per cent. This effect is even more important for $n_{\rm
    s}$ as it increases the error-bars by more than 30 per cent. As expected,
  the cosmological parameters are then less biased. Nevertheless,
  $n_{\rm s}$ remains significantly affected as the bias is still more
  than 3 times its expected accuracy. The bias on $\sigma_8$ is also
  significant. Consequently, a more appropriate analysis
would rather consist in fitting the CMB data with a coherent model
accounting for both the primary and the residual signal with
  full dependency on the cosmological parameters.  For that purpose,
one would need to use the cluster selection function, the
understanding of which is rather complex (instrumental effects,
foregrounds, limits due to the component separation techniques and
cluster extraction methods, etc). In that case, a better solution
would be rather to use the total signal including the primary CMB and
the secondary anisotropies to determine the cosmological parameters in
a coherent way, that is taking into account the full cosmological
dependence of the secondary signal. The latter is crucial, as it was
shown \citep[][]{Marian06} that modelling the SZ contribution by a
fixed-shape power spectrum with a varying amplitude $A_{\rm SZ}$ is
not sufficient and biases the cosmological parameters.

\section{Conclusions}

In this study, we develop an analytical method to calculate the
biases on the cosmological parameters. Our method applies to any
contamination provided the primary signal and the contamination are
additive. Additionally, it is an exact derivation to the
  second order with the advantage of being applicable even
when the contaminant dominates over the primary signal. The next
generation of CMB experiments will measure, with a high sensitivity,
the signal at small angular scales where secondary contributions
intervene. We apply our method to the case of a contribution from
undetected thermal SZ clusters to the primary CMB signal (assuming no
contribution from foregrounds or point sources).

For illustration, we take the characteristics (noise, beam size) of
\emph{Planck} and compute the residual SZ signal from the undetected
clusters assuming all clusters above 3 or 5$\sigma_Y$ are detected
simultaneously in the channels 100, 143 and 353~GHz. The residual SZ
signal contributes more than 10 per cent of the total signal at multipoles
higher than 1000. The higher the SZ cluster detection threshold, the
higher the contamination.

We perform a bias estimation simultaneously on six cosmological
parameters ($\Omega_\Lambda$, $\Omega_{\rm b}$, $H_0$, $n_{\rm s}$,
$\sigma_8$ and $\tau$) using temperature and polarisation anisotropy
TT, TE and EE power spectra. This quantifies the effect of fitting the
data, that include a residual contribution, with a model that ignores
it.  We then compare the biases to the expected 1$\sigma$ errors on
each parameter. We find that the biases induced by the thermal SZ
residual signal on $\Omega_\Lambda$ and $H_0$ are
negligible. At most, they are of the order of 0.08
  and 0.6 in units of the 1$\sigma$ error on the parameters. On the
contrary, the determination of $\Omega_{\rm b}$, $n_{\rm s}$
  and $\tau$ is significantly altered by the residual SZ signal. The
  biases are 2.4, 6. and 10.4$\sigma$ respectively. This is easily
  understood as they are the most sensitive parameters to an excess of
  power at small scales and, moreover, they are degenerate.

We point out the importance of taking into account the SZ residuals in
the analysis of the small scale high sensitivity CMB
data. The SZ residual power spectrum depends on the cosmological
parameters and on the cluster selection function. A joint analysis of
primary and secondary CMB signal will provide additional constraints
on the cosmological parameters and thus reduce the biases
arising from the SZ residual power excess at high
  multipoles. A coherent analysis, including full cosmological
  parameters dependency of the primary and secondary signal, allow one
  to use the whole range of multipoles, including the highest ones.

\section*{Acknowledgments}
The authors thank the referee D. Scott for useful comments and
  suggestions on the manuscript. We further thank
  A. Challinor, H. Dole, G. Lagache, and B. M. Sch\"afer for useful
  discussions.  We aknowledge the use of the CAMB package.
\label{lastpage}

\bibliographystyle{mn2e}
\bibliography{TaburetN_oct08}
\end{document}